\documentclass[review]{elsarticle}

\usepackage{lineno,hyperref,setspace,caption,multirow,booktabs,rotating,color,ulem}
\modulolinenumbers[5]

\newcommand{\tb}[1]{\textcolor{black}{#1}}
\newcommand{\gm}[1]{\textcolor{black}{#1}}

\journal{Chemical Physics Letters}

\begin{document}

\begin{frontmatter}

\title{\tb{Assessment of \gm{the} Performance of} DFT Functionals Using Off-Diagonal Hypervirial Relationships}

\author{Francesco F. Summa, Guglielmo Monaco, Riccardo Zanasi\corref{mycorrespondingauthor}}
\address{Dipartimento di Chimica e Biologia ``A. Zambelli'', Universit\`{a} degli studi di Salerno, via Giovanni Paolo II 132, Fisciano 84084, SA, Italy}
\author{Lazzeretti Paolo}
\address{Istituto di Struttura della Materia, CNR, via del Fosso del Cavaliere 100, 00133 Roma, Italy}


\cortext[mycorrespondingauthor]{Corresponding author: Riccardo Zanasi, \textit{Email Address:} rzanasi@unisa.it}


\begin{abstract}
	Off-diagonal hypervirial relationships, combined with quantum mechanical sum rules of charge-current conservation, offer a way for testing electronic excited-state transition energies and moments, which does not need any external reference. A number of fundamental relations were recast into absolute deviations from zero, which have been used to  \tb{assess the performance of} some popular DFT functionals. Extended TD-DFT calculations have been  \tb{carried out} for a pool of molecules chosen to the purpose, adopting a large basis set to ensure high quality results.  \tb{A} \tb{general} agreement with previous benchmarks \tb{is observed}.
\end{abstract}

\begin{keyword}
\texttt{Transition Moments  \sep TRK Sum Rules \sep Magnetizability Traslational Invariance \sep Charge-Current Conservation \sep Electric Dipole Polarizability \sep Optical Activity }
\end{keyword}

\end{frontmatter}


\section{Introduction}
Owing to the impressive use of density functional theory (DFT) in last years for quantum chemistry calculations on many-electron systems, the quality assessment of DFT functionals has become a rather popular and important activity \cite{medvedev_density_2017,jacquemin_extensive_2009,goerigk_thorough_2011,ali_td-dft_2020,lehtola_benchmarking_2020,sen_benchmarking_2020}.  Unlike what has been done so far, here we present  \tb{an assessment of \gm{the} performance} of DFT functionals on the basis of  excited-states transition moments and energies,  \tb{which could reveal useful in the choice of functionals for the calculation of second-order molecular properties arising from the interaction with radiation}. 

DFT is usually considered as a ground-state theory. However, as proven by Hohenberg and Kohn (for systems with a nondegenerate ground state) the ground-state electron probability density $\rho_{0}(\mathbf{r})$ determines the external potential and the number of electrons \cite{hohenberg_inhomogeneous_1964,parr_density-functional_1989}. Once the external potential and the number of electrons are specified, the electronic wave functions and allowed energies of the molecule are determined as the solutions of the electronic Schr{\"o}dinger equation. Hence, ground-state wave function and energy as well as all excited-state wave functions and energies are determined by $\rho_{0}(\mathbf{r})$ \cite{koch_chemists_2008}. However, the extraction of such information remained unsolved until the presentation of the formally exact formulation of DFT for time-dependent (TD) systems given by Runge and Gross \cite{runge_density-functional_1984}.  Actually, electronic excited-states reveal themselves from the interaction with radiation, and the development of the linear response TD-DFT via approximated functionals has provided direct access to the excited-state information searched for \cite{chong_time-dependent_1995,stratmann_efficient_1998,casida_progress_2012,adamo_calculations_2013}.

The accuracy of a calculation of molecular properties depending on the electronic response to a perturbation could be established via direct comparison with corresponding experimental data. However, this is not always an easy task to accomplish because the experimental data can be affected by effects, such as intermolecular interactions, which can be poorly described by the theoretical model, and the computations themselves can carry an uncertainty due to finiteness of the basis set. As far as it concerns the assessment of DFT predictions, the comparison with higher level calculations is often employed \cite{miura_assessment_2007,sarkar_benchmarking_2021}, a practice which should require the adoption of  \tb{difficult} reference calculations for obtaining a representative and robust grading of the methods. The fundamental need for instruments assessing the quality of a theoretical prediction \textit{a priori}, i.e., without external references, is evident. For testing excited-states information resulting from a TD-DFT calculation, these instruments can be supplied by some internal quantum mechanical conditions, such as the fulfillment of the off-diagonal hypervirial relationships, Thomas-Reiche-Kuhn sum rules for the oscillatory strengths \cite{thomas_uber_1925,reiche_ber_1926,kuhn_ber_1925} and other sum rules for the charge-current conservation \cite{sambe_properties_1973}, which are equivalent to the translational invariance of the magnetizability \cite{arrighini_magnetic_1968}. 

In the next section we will focus on the definition of a number of absolute deviations that do not need any external reference. Then, results will be presented for a selection of functionals chosen among the most popular for the calculation of molecular response properties. Testing molecules of moderate size have been chosen to allow accurate calculations of the full set of TD-DFT excited states required for the linear response, adopting a fairly large set of basis functions.

\section{Method}

Let's consider a molecule with $n$ electrons of mass $m_\textrm{e}$, charge $-e$, positions $\mathbf{r}_{k}$, canonical and angular momenta $\hat{\mathbf{p}}_{k}$ and $\hat{\mathbf{l}}_k=\mathbf{r}_k\times\hat{\mathbf{p}}_k$, respectively. Let's consider also  the origin of the coordinate system in the center of positive charges. Eigenvalues of the Born-Oppenheimer Hamiltonian $\hat{H}_0$ are $E_0$ and $E_j$, which are energies of the stationary states $\Psi_0$ and $\Psi_j$, respectively (subscript 0 denoting the ground reference state and $j$ any excited state). Capital letters are used to denote total electron operators:
\begin{equation}
	\hat{R}_{\alpha}=\sum_k{r}_{k\alpha}, \qquad \hat{P}_{\alpha}=\sum_k\hat{p}_{k\alpha}, \qquad \hat{L}_{\alpha}=\sum_k\hat{l}_{k\alpha},
	\label{Operators}
\end{equation}
where
\begin{equation}
	\hat{P}_\alpha=\frac{\textrm{i}m_\textrm{e}}{\hbar}\left[\hat{H}_0,\hat{R}_\alpha\right].
	\label{commut}
\end{equation}

Transition moments of $\hat{R}_{\alpha}$ and $\hat{P}_{\alpha}$ from ground to excited stationary states are related by the off-diagonal hypervirial relation 
\begin{equation}
	\left\langle 0\left|\hat{R}_{\alpha}\right| j\right\rangle=
	\frac{\textrm{i}}{m_\textrm{e}} \omega_{j0}^{-1}
	\left\langle 0\left|\hat{P}_{\alpha}\right| j\right\rangle,
	\label{offHR}
\end{equation}
where $\omega_{j0}=\frac{1}{\hbar}(E_j-E_0)$,  which is fulfilled if the state functions are \textit{exact} eigenfunctions of a model Hamiltonian and satisfy the hypervirial theorem for the position operator \cite{epstein_variation_1974}. 
The  \tb{TD Hartree-Fock  (TD-HF)} method fulfils eq.(\ref{offHR}) in the limit of a complete basis set calculation \cite{la_paglia_theoretical_1966,harris_oscillator_1969} \tb{and it is generally assumed that TD-DFT does the same, see for example Ref. \cite{sarkar_benchmarking_2021}. Indeed, there is a very well know connection among current density conservation, gauge invariance of magnetic properties and hypervirial relationship as proven by Ghosh and Dhara for Kohn-Sham-like approach \cite{ghosh_density-functional_1988,ding_gauge_2011}. Excitation energies, as well as transition matrix elements of the position, linear and angular momentum operators, are main results of any TD calculation and it would appear inappropriate to have large deviations among the oscillator and rotational strengths determined in the so-called length, velocity or mixed gauges (the latter applies only to oscillator strengths, of course).}

The Thomas-Reiche-Kuhn (TRK) \cite{thomas_uber_1925,reiche_ber_1926,kuhn_ber_1925} sum rule for the position operator in tensor notation is \cite{lazzeretti_quantum-mechanical_1985,Lazzeretti_AdvChemPhys} 
\begin{equation}
	(\hat{R}_{\alpha}, \hat{R}_{\beta})_{1}=
	\frac{m_\textrm{e}}{\hbar}
	{\sum_{j}}^{\prime}  2 \omega_{j0}
	\Re \left(\left\langle 0\left| \hat{R}_\alpha \right| j\right\rangle
	\left\langle j \left| \hat{R}_\beta \right| 0\right\rangle\right)
	=n\delta_{\alpha\beta}.
	\label{TRK_RR}
\end{equation}
Using eq.(\ref{offHR}) one gets alternative mixed velocity-length and velocity-velocity formulations:
\begin{equation}
	(\hat{P}_{\alpha}, \hat{R}_{\beta})_{0}=
	\frac{1}{\hbar}
	{\sum_{j}}^{\prime} 2
	\Im \left(\left\langle 0\left|\hat{P}_{\alpha}\right| j\right\rangle
	\left\langle j \left| \hat{R}_\beta \right| 0\right\rangle\right)
	=n\delta_{\alpha\beta},
	\label{TRK_PR}
\end{equation}
\begin{equation}
	(\hat{P}_{\alpha}, \hat{P}_{\beta})_{-1}=
   \frac{1}{m_\textrm{e}\hbar}
   {\sum_{j}}^{\prime}  \frac{2}{\omega_{j0}}
	\Re \left(\left\langle 0\left|\hat{P}_{\alpha}\right| j\right\rangle\left\langle j\left|\hat{P}_{\beta}\right| 0\right\rangle\right)
	=n\delta_{\alpha\beta},
	\label{TRK_PP}
\end{equation}
which can be briefly summarized as
\begin{equation}
	n\delta_{\alpha\beta}=(\hat{P}_\alpha,\hat{P}_\beta)_{-1}=(\hat{P}_\alpha,\hat{R}_\beta)_{0}=(\hat{R}_\alpha,\hat{R}_\beta)_{1}.
	\label{TRK}
\end{equation}

Additional sum rules for the translational invariance of the magnetizabiliy tensor, which are intimately connected to the charge-current conservation condition of Sambe \cite{sambe_properties_1973}, are \cite{arrighini_magnetic_1968,lazzeretti_quantum-mechanical_1985,Lazzeretti_AdvChemPhys} 
\begin{eqnarray}
	\left\langle 0 \left| \hat{R}_{\beta}\right| 0 \right\rangle\epsilon_{\gamma\alpha\beta}&=&\left(\hat{P}_{\gamma}, \hat{L}_{\alpha}\right)_{-1}=
	-\left(\hat{P}_{\alpha}, \hat{L}_{\gamma}\right)_{-1} \nonumber \\
	&=&\left(\hat{R}_{\gamma}, \hat{L}_{\alpha}\right)_{0}=-\left(\hat{L}_{\gamma}, \hat{R}_{\alpha}\right)_{0},
	\label{TIM}
\end{eqnarray}
where $\epsilon_{\gamma\alpha\beta}$ is the Levi-Civita skew-symmetric tensor and
\begin{equation}
	(\hat{R}_{\gamma}, \hat{L}_{\alpha})_{0}=
	-\frac{1}{\hbar}
	{\sum_{j}}^{\prime} 2
	\Im \left(\left\langle 0\left|\hat{R}_{\gamma}\right| j\right\rangle\left\langle j\left|\hat{L}_{\alpha}\right| 0\right\rangle\right),
	\label{TIM_RL}
\end{equation}
\begin{equation}
	(\hat{P}_{\gamma}, \hat{L}_{\alpha})_{-1}=
	\frac{1}{m_\textrm{e}\hbar}
	{\sum_{j}}^{\prime} \frac{2}{\omega_{j0}}
	\Re \left(\left\langle 0\left|\hat{P}_{\gamma}\right| j\right\rangle\left\langle j\left|\hat{L}_{\alpha}\right| 0\right\rangle\right).
	\label{TIM_PL}
\end{equation}

Allowing for the off-diagonal hypervirial relation (\ref{offHR}), the frequency dependent electric dipole polarizability can be written in three formalisms \cite{Lazzeretti_AdvChemPhys,lazzeretti_electric_1986}, in a way similar to that used for TRK sum rules (\ref{TRK_RR})-(\ref{TRK_PP}). They are 
\begin{equation}
	\alpha_{\alpha\beta}^{RR}(\omega)=
	\frac{e^2}{\hbar}
	{\sum_{j}}^{\prime}
	\frac{2\omega_{j0}}{\omega_{j0}^2-\omega^2}
	\Re \left(\left\langle 0\left| \hat{R}_\alpha \right| j\right\rangle
	\left\langle j \left| \hat{R}_\beta \right| 0\right\rangle\right),
	\label{alphaRR}
\end{equation}
\begin{equation}
	\alpha_{\alpha\beta}^{PR}(\omega)=
	\frac{e^2}{m_\textrm{e}\hbar}
	{\sum_{j}}^{\prime} \frac{2}{\omega_{j0}^2-\omega^2}
	\Im \left(\left\langle 0\left| \hat{P}_\alpha \right| j\right\rangle
	\left\langle j \left| \hat{R}_\beta \right| 0\right\rangle\right),
	\label{alphaPR}
\end{equation}
\begin{equation}
	\alpha_{\alpha\beta}^{PP}(\omega)=
	\frac{e^2}{m_\textrm{e}^2\hbar}
	{\sum_{j}}^{\prime}\frac{2}{\omega_{j0}(\omega_{j0}^2-\omega^2)}
	\Re \left(\left\langle 0\left| \hat{P}_\alpha \right| j\right\rangle
	\left\langle j \left| \hat{P}_\beta \right| 0\right\rangle\right),
	\label{alphaPP}
\end{equation}
whilst the frequency dependent optical activity tensor in two formalisms is \cite{Lazzeretti_AdvChemPhys,lazzeretti_electric_1986}
\begin{equation}
	\hat{\kappa}_{\alpha\beta}^{RL}(\omega)=
	-\frac{e^2}{2m_\textrm{e}\hbar}
	{\sum_{j}}^{\prime}
	\frac{2}{\omega_{j0}^2-\omega^2}
	\Im \left(\left\langle 0\left| \hat{R}_\alpha \right| j\right\rangle
	\left\langle j \left| \hat{L}_\beta \right| 0\right\rangle\right),
	\label{kappaRL}
\end{equation}
\begin{equation}
	\hat{\kappa}_{\alpha\beta}^{PL}(\omega)=
	\frac{e^2}{2m_\textrm{e}^2\hbar}
	{\sum_{j}}^{\prime}
	\frac{2}{\omega_{j0}(\omega_{j0}^2-\omega^2)}
	\Re \left(\left\langle 0\left| \hat{P}_\alpha \right| j\right\rangle
	\left\langle j \left| \hat{L}_\beta \right| 0\right\rangle\right).
	\label{kappaPL}
\end{equation}

A direct measure of the ability of any density functional to provide a good description of excited states can be obtained adopting self-consistent absolute deviations defined by means of the above equations without resorting to any external references. Considering first eq. (\ref{offHR}), we define the absolute deviation for the off-diagonal hypervirial relation, which collects at the same time all vector components and computed electronic transitions, as (atomic units)
\begin{equation}
	\textrm{AD}_{\footnotesize\textrm{HV}}=\sum_\alpha \sum_{j} 
	\left| 	
	\left\langle 0\left|\hat{R}_{\alpha}\right| j\right\rangle -
	\frac{ \left\langle 0\left|{\nabla}_{\alpha}\right| j\right\rangle  }{E_j-E_0} 
	\right| ,
	\label{adhv}
\end{equation}
which should be as small as possible for a precise calculation. Moreover,
from eq. (\ref{TRK}), we define the three following absolute deviations testing the TRK sum rules
\begin{equation}
	\textrm{AD}_{\footnotesize\textrm{RR}} = \sum_\alpha\sum_\beta \left|
	(R_\alpha,R_\beta)_{1}-n\delta_{\alpha\beta}
	\right| ,
	\label{adrr}
\end{equation}
\begin{equation}
	\textrm{AD}_{\footnotesize\textrm{PR}} = \sum_\alpha\sum_\beta \left|
	(P_\alpha,R_\beta)_{0}-n\delta_{\alpha\beta}
	\right| ,
	\label{adpr}
\end{equation}
\begin{equation}
		\textrm{AD}_{\footnotesize\textrm{PP}} = \sum_\alpha\sum_\beta \left|
		(P_\alpha,P_\beta)_{-1}-n\delta_{\alpha\beta}
		\right| .
		\label{adpp}
\end{equation}
Once again they should be as small as possible for accurate calculations.
Condition (\ref{TIM}) provides two further absolute deviations
\begin{equation}
	   \textrm{AD}_{\footnotesize\textrm{RL}} = \sum_\gamma \sum_\alpha 
	   \left|
	   \left(R_{\gamma}, L_{\alpha}\right)_{-1}-\epsilon_{\gamma\alpha\beta}
		\left\langle 0 \left| \hat{R}_{\beta}\right| 0 \right\rangle\right| ,
		\label{adrl}
\end{equation}
\begin{equation}
	\textrm{AD}_{\footnotesize\textrm{PL}} = \sum_\gamma \sum_\alpha 
	\left|
	\left(P_{\gamma}, L_{\alpha}\right)_{-1}-\epsilon_{\gamma\alpha\beta}
	\left\langle 0 \left| \hat{R}_{\beta}\right| 0 \right\rangle\right| .
	\label{adpl}
\end{equation}

A slightly different way to assess the off-diagonal hyperviral relation (\ref{offHR}) is to check the equivalence of the computed electric dipole polarizability in the various formalisms provided by eqs. (\ref{alphaRR})-(\ref{alphaPP}). This can be done defining the following absolute deviation
\begin{equation}
	\textrm{AD}_{\alpha}= \sum_\beta \sum_\gamma 
	\left(
	\left| \alpha_{\beta\gamma}^{RR}-\alpha_{\beta\gamma}^{PR} \right|+
	\left| \alpha_{\beta\gamma}^{RR}-\alpha_{\beta\gamma}^{PP} \right|+
	\left| \alpha_{\beta\gamma}^{PR}-\alpha_{\beta\gamma}^{PP} \right|
	\right) .
	\label{adalpha}
\end{equation}
In case of chiral molecules also the optical activity tensor calculated using eqs. (\ref{kappaRL}) and (\ref{kappaPL}) can be used to the purpose by means of
\begin{equation}
	\textrm{AD}_{\hat{\kappa}}= \sum_\beta \sum_\gamma 
	\left| 
	\hat{\kappa}_{\beta\gamma}^{RL}-
	\hat{\kappa}_{\beta\gamma}^{PL} \right| .
	\label{adkappa}
\end{equation}
Of course, also $\textrm{AD}_{\alpha}$ and $\textrm{AD}_{\hat{\kappa}}$ should be as small as possible. 

\section{Calculation details}
 \tb{Nineteen} DFT functionals have been tested on the basis of the absolute deviations defined in the previous section. They are: 
APFD \cite{austin_density_2012}, 
B3LYP \cite{becke_densityfunctional_1993,lee_development_1988,miehlich_results_1989}, 
B3P86 \cite{becke_densityfunctional_1993,perdew_density-functional_1986}, 
B3PW91 \cite{becke_densityfunctional_1993,perdew_atoms_1992,perdew_erratum_1993,perdew_generalized_1996-1}, 
B97-2 \cite{wilson_hybrid_2001}, 
B97D3 \cite{grimme_effect_2011}, 
B97D \cite{grimme_semiempirical_2006}, 
B98 \cite{schmider_optimized_1998}, 
BLYP \cite{becke_density-functional_1988,lee_development_1988,miehlich_results_1989}, 
CAM-B3LYP \cite{yanai_new_2004}, 
SVWN \cite{slater_quantum_1974,vosko_accurate_1980} (indicated as LSDA for our purposes), 
MN12-SX \cite{peverati_screened-exchange_2012}, 
M06 \cite{zhao_m06_2008}, 
M06-2X \cite{zhao_m06_2008}, 
PBEP86 \cite{perdew_generalized_1996,perdew_generalized_1997,perdew_density-functional_1986},
\tb{BHandHLYP} \cite{becke_new_1993}, and
\tb{LC-BLYP, LC-B97D, LC-BP86} \cite{iikura_long-range_2001}.

These functionals have been chosen considering the analysis by Medvedev et al \cite{medvedev_density_2017}, selecting some commonly used among those \tb{``yielding the best densities''} (APFD, B3PW91, B98, \tb{BHandHLYP}, B3P86, B972 and B3LYP), other among those \tb{``yielding the worst densities''} \cite{nota1} (SVWN, M06-2X, M06 and MN12-SX) and  some other providing densities of intermediate quality (CAM-B3LYP, PBEP86 and BLYP). Two more functionals (B97D and B97D3) came from our previous experiences, \gm{and the three long-range corrected (LC) functionals have been included because long-range correction is a relevant correction in DFT as it helps providing good excitation energies \cite{hirao_charge-transfer_2020}.}

Calculations have been carried out for a set of molecules selected according to the following criteria: 
\begin{itemize}
	\item  \tb{the molecular size should not limit the calculation accuracy, which must be as large as possible in order to obtain reliable estimates of deviations (\ref{adhv})-(\ref{adkappa})}; 	
	\item all having the same number of electrons for a better comparison of the TRK sum rules in the different formalisms; 
	\item all having \tb{the possibility to be fixed in} a chiral conformation with a non vanishing dipole moment. 
\end{itemize}
Accordingly, we have considered $N=6$ isoelectronic molecules: 
CH$_3$-CH$_3$, 
CH$_3$-NH$_2$, 
CH$_3$-OH, 
NH$_2$-NH$_2$, 
NH$_2$-OH, 
HO-OH, 
with $n=18$ electrons. \tb{Taking into account the very large basis set employed for the calculations, see later, the chosen molecules fulfill the above criteria and appear well suited for the purposes of the present investigation.} 

For each functional F and for each molecule M, absolute deviations will be indicated as $\textrm{AD}^{\footnotesize\textrm{F}}_{\footnotesize\textrm{X}}(\textrm{M})$, where X stands for one of the testing conditions among: HV, RR, PR, PP, RL, PL, $\alpha$, and $\hat{\kappa}$. Mean absolute deviations have been computed for each DFT functional and each testing condition averaging the results obtained for the $N$ molecules and normalizing to the HF results
\begin{eqnarray}
	\textrm{MAD}^{\footnotesize\textrm{F}}_{\footnotesize\textrm{X}}&=&
	\frac{
	{\frac{1}{N}\sum_{\footnotesize\textrm{M}}
		\textrm{AD}^{\footnotesize\textrm{F}}_{\footnotesize\textrm{X}}(\textrm{M})}}{\textrm{MAD}^{\footnotesize\textrm{HF}}_{\footnotesize\textrm{X}}} \\	\textrm{MAD}^{\footnotesize\textrm{HF}}_{\footnotesize\textrm{X}}&=&{\frac{1}{N}\sum_{\footnotesize\textrm{M}}
		\textrm{AD}^{\footnotesize\textrm{HF}}_{\footnotesize\textrm{X}}(\textrm{M})} . \nonumber
	\label{MADs}
\end{eqnarray}

Molecular geometries have been determined as follows. For each functional, internal coordinates have been optimized employing the largest pcSseg-4 of the Jensen's set \cite{jensen_segmented_2015}, freezing the internal torsional angle of each molecule at the values reported in Table \ref{diedri}. A total of  \tb{120} \tb{(19 functionals + HF times 6 molecules)} geometries have been determined, which are reported in the Supporting Material, see Tables S1-\tb{S20}. Torsional angles have been determined separately, taking the value that maximizes the largest component of the optical activity tensor of each molecule, along a rotation around the single bond connecting the two heaviest atoms. This latter determination has been done adopting the much smaller pcSseg-2 basis set \cite{jensen_segmented_2015} and using the B97-2 functional \cite{wilson_hybrid_2001} for all molecules.

\begin{table}[htbp]
	\caption{Torsional angles frozen during geometry optimization and number of calculated transitions NTRS for the molecules here considered.}
	\centering
	\begin{tabular}{ccccccc}
		\hline
		&
		CH$_3$-CH$_3$ & 
		CH$_3$-NH$_2$ &
		CH$_3$-OH & 
		NH$_2$-NH$_2$ & 
		NH$_2$-OH & 
		HO-OH \\
		\hline
		$\theta$ &
		270$^\circ$ & 90$^\circ$  & 210$^\circ$  & 159$^\circ$  & 180$^\circ$  & 120$^{\circ\dagger}$ \\
		\hline
		NTRS & 6624 & 5958 & 5265 & 5283 & 4599 & 3906 \\
		\hline
	\end{tabular}
{\footnotesize $^\dagger$ Average value, since the molecular geometry has been fully optimized every time.}
	\label{diedri}
\end{table}

TD-DFT calculations of transition matrix elements  
$\left\langle 0\left|\hat{R}_{\alpha}\right| j\right\rangle$,
$\left\langle 0\left|\hat{P}_{\alpha}\right| j\right\rangle$, 
$\left\langle 0\left|\hat{L}_{\alpha}\right| j\right\rangle$ and 
transition energies 
$(E_j-E_0)$, for $j=1,2,\dots,\textrm{NTRS}$, have been performed adopting the same large pcSseg-4 basis set of the geometry optimization, by means of the Gaussian 16 program package \cite{g16}, using the TD=(full,sos) keyword. The pcSseg-4 basis set provides high quality molecular properties nearly coincident with the basis set limit, as documented previously \cite{monaco_atomic_2020}.  \tb{Despite the small size of the molecules, a fairly large amount of computer resources become necessary to accomplish the full set of calculations, much more than what is normally available on a PC.}

Both the electric dipole polarizability and the optical activity tensors have been evaluated for the static case $\omega=0$. 

\section{Results and discussion}
The full set of results for all molecules is collected in the Supporting Material. The discussion here is focused on a number of mean absolute deviations, which represent the essential result of the analysis. 

Let us first comment on that none of the DFT functionals here considered provide any AD's\tb{, except 6 over a total of 912 examined,} less than those obtained at the HF level.\cite{nota2} \tb{Anyway, this} result  \gm{can}not be regarded as an alleged superiority of the HF method \gm{(although this happens for some properties\cite{sen_benchmarking_2020})} because it depends on the fulfillment of the off-diagonal hypervirial relation (\ref{offHR}) for the \tb{TD-}HF state functions in the \gm{complete} basis set limit \cite{harris_oscillator_1969}. We exploit this in eq.(\ref{MADs}) to put all the results on the same scale in order to make data aggregations.

\begin{figure}[htbp]
	\centering
	\caption{Absolute deviations calculated according to eq. \ref{adhv} and averaged over the molecular set.}
	\includegraphics[scale=0.9]{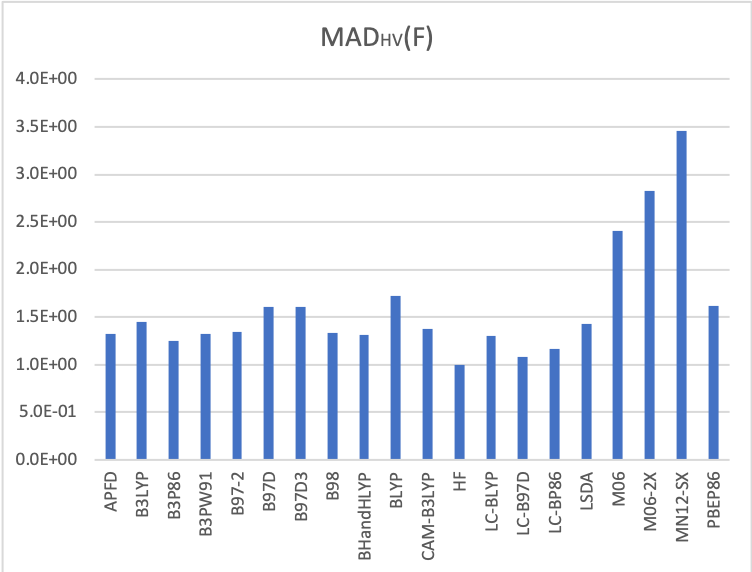}
	\label{madfhv}
\end{figure}

Mean absolute deviations for the off-diagonal hypervirial relation, i.e.,  $\textrm{MAD}^{\footnotesize\textrm{F}}_{\footnotesize\textrm{HV}}$, calculated averaging the data in Table  \tb{S21}, are compared in Fig. \ref{madfhv}. As can be observed, the \tb{LC-functionals,} B3P86 \tb{and BHandHLYP} provide the smallest deviation\tb{s}, whilst \gm{the Minnesota functionals} M06, M06-2X, and MN12-SX give the highest deviations.  \tb{However,} a substantial agreement with previous benchmarks \cite{medvedev_density_2017,jacquemin_extensive_2009,goerigk_thorough_2011,lehtola_benchmarking_2020} can be observed, even if we found B3P86 \tb{and BHandHLYP} slightly better than APFD. 

\begin{figure}[htbp]
	\centering
	\caption{Normalized sum of the mean absolute deviations relative to the TRK sum rule in the various formalisms. M06, M06-2X, and MN12-SX reach a MAD as high as 14, 36, and 34 with respect to the HF, respectively.}	
	\includegraphics[scale=0.9]{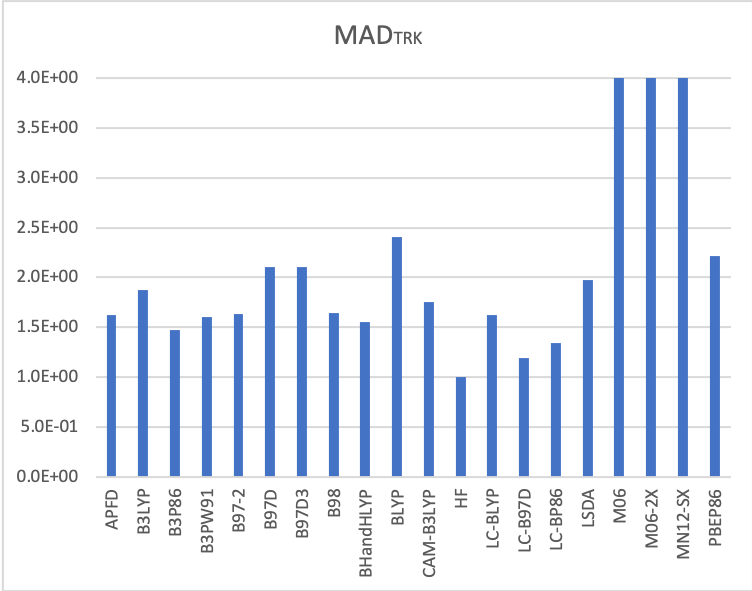}
	\label{madtrk}
\end{figure}

Considering the TRK sum rule, we notice some quite large deviations for  \gm{the Minnesota} functionals in connection with RR and PP formalisms, see Tables  \tb{S22} and  \tb{S24}, which are not shown for the PR formalism, see Table  \tb{S23}. This behavior is interesting since it reveals a particular sensitivity of these functionals with respect to transition energies, see eqs. (\ref{TRK_RR})-(\ref{TRK_PP}). All other functionals provide a consistent picture, which can be summarized into the data aggregation given by the normalized sum of the three mean absolute deviations
$ \frac{1}{3}\left(\textrm{MAD}^{\footnotesize\textrm{F}}_{\footnotesize\textrm{RR}} + \textrm{MAD}^{\footnotesize\textrm{F}}_{\footnotesize\textrm{PR}} +
\textrm{MAD}^{\footnotesize\textrm{F}}_{\footnotesize\textrm{PP}}\right) $
shown in Fig. \ref{madtrk}.

\begin{figure}[htbp]
	\centering
	\caption{Normalized sum of the mean absolute deviations relative to the translation invariance of magnetizability in the two formalisms. M06, M06-2X, and MN12-SX reach a MAD as high as 5, 3, and 8 with respect to the HF, respectively.}	
	\includegraphics[scale=0.9]{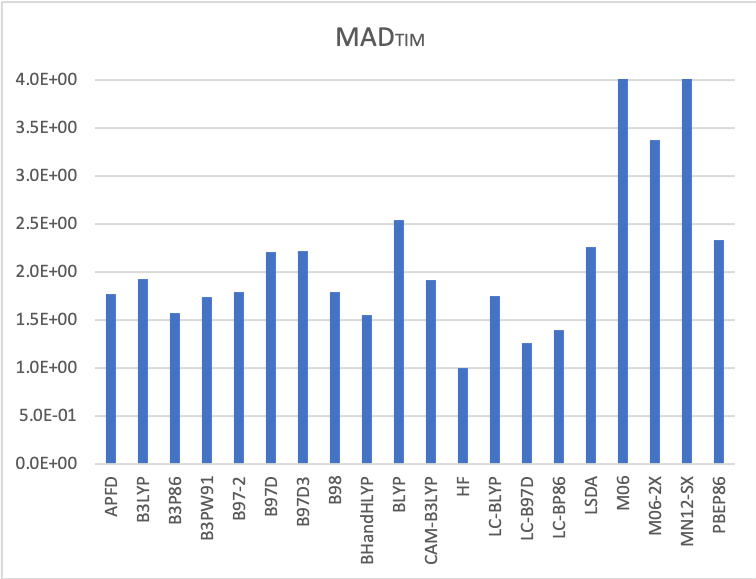}
	\label{madtim}
\end{figure}

A similar behavior can be observed also in the case of the sum rule for the translational invariance of magnetizability. Once again, albeit to a lesser extent, the PL formalism shows large deviations for M06, M06-2X, and MN12-SX functionals, see Table \tb{S26}, which are not shown by the RL formalism, see Table  \tb{S25}, confirming the sensitivity of these functionals to the transition energies that appear in eq.(\ref{TIM_PL}). The aggregated data given as the normalized sum $\frac{1}{2}\left(\textrm{MAD}^{\footnotesize\textrm{F}}_{\footnotesize\textrm{RL}} + \textrm{MAD}^{\footnotesize\textrm{F}}_{\footnotesize\textrm{PL}} \right)$
shown in Fig. \ref{madtim}, confirms the functional  \tb{performance} of the hypervirial and TRK analysis.

The absolute deviations that measure the equivalence of the static electric dipole polarizability and optical activity tensor in the various formalisms, reported in Tables  \tb{S27} and  \tb{S28}, and relative Figs. S7 and S8, do not change the picture. 

\begin{figure}[htbp]
	\centering
	\caption{Mean absolute deviations for all criteria.}	
	\includegraphics[scale=0.7]{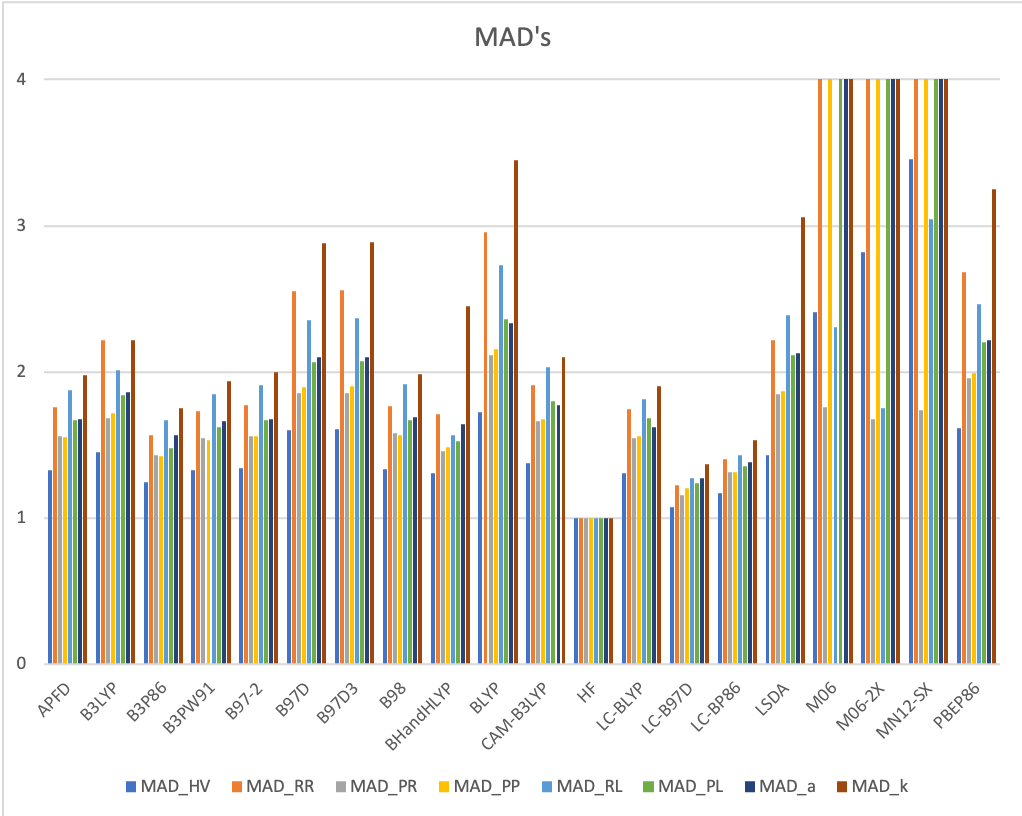}
	\label{madall}
\end{figure}
 Therefore, to best appreciate the uniformity of the results, we have put together all the MAD's for the eight criteria in the same histogram chart shown in Fig. \ref{madall}. As it can be seen, the functional \tb{performance} turns out to be exactly the same regardless of the MAD. Encouraged by this rather nice result, we have decided to work out a unique index for \tb{ranking} the functionals here considered, summing all together the mean absolute deviations as 
 \begin{equation}
 	\textrm{I}=\frac{1}{8}\sum_\textrm{X} \textrm{MAD}^{\footnotesize\textrm{F}}_{\footnotesize\textrm{X}}.
 	\label{index}
 \end{equation}
The  \tb{index} is shown in Table \ref{totalMAD} and the relative histogram chart constitutes the graphical abstract of the paper.

\begin{table}
	\centering
	\caption{Functional ranking based on the grading index values calculated via eq. (\ref{index}). \gm{The rung of Jacob's ladder\cite{perdew_2001} is also indicated as LDA (Local Density Approximation), GGA (Generalised Gradient Approximation), mGGA (meta-GGA), hGGA (hybrid-GGA).}}
	\begin{tabular}{|l|r|l|c|c|}
		\hline
		Functional & Rung & Index & MBSPL$^a$ & LDFS$^b$ \\ \hline
		HF & Ab Initio& 1 & & \\  \hline
		LC-B97D &hGGA& 1.228 & & \\ \hline
		LC-BP86 &hGGA& 1.365 & & \\ \hline
		B3P86 &hGGA& 1.517 &10 &  \\ \hline
		BHandHLYP	&hGGA& 1.643 & 6 & 1\\ \hline
		LC-BLYP &hGGA& 1.648 & & \\ \hline
		B3PW91 &hGGA& 1.651 & 3 & \\ \hline
		APFD &hGGA& 1.676 & 2 & \\ \hline
		B97-2 &hGGA& 1.687 &17 & 29\\ \hline
		B98 &hGGA& 1.690 & 5 & \\ \hline
		CAM-B3LYP &hGGA& 1.793 & 48 & \\ \hline
		B3LYP &hGGA& 1.876 & 55 & 16\\ \hline
		LSDA &LDA& 2.133 & 109 & 42 \\ \hline
		B97D &GGA& 2.164 & & \\ \hline
		B97D3 &GGA& 2.170 & & \\ \hline
		PBEP86 &GGA& 2.298 & 70 & \\ \hline
		BLYP &GGA& 2.478 & 92 & 31 \\ \hline
		M06 &hmGGA& 9.921 & 121 & 51 \\ \hline
		MN12-SX &hmGGA& 16.579 & 128 & \\ \hline
		M06-2X &hmGGA& 18.770 & 114 & 45 \\ \hline
	\end{tabular}

$^a$ Electron density-based normalized error rank over 128 functionals from Medvedev et al. \cite{medvedev_density_2017}. 

$^b$ Magnetizability-based mean absolute deviation rank over 51 functionals from Lehtola et al. \cite{lehtola_benchmarking_2020}.
	\label{totalMAD}
\end{table}

According to this result, the \tb{LC-B97D, LC-BP86,} B3P86 \tb{and BHandHLYP functionals} turn out to be \tb{good} choice\tb{s} among the functionals here considered for the calculation of the linear response properties \tb{to external electromagnetic perturbations. A fairly good agreement with previous analysis, reported by Medvedev et al. \cite{medvedev_density_2017} and Lehtola et al. \cite{lehtola_benchmarking_2020}, can be appreciated, as we have facilitated adding more columns to Table \ref{totalMAD}, where the rank attributed by these two different \gm{analyses} is reported. Since Lethola et al. have considered magnetizability as the benchmarking property, which is strictly connected with our RL and PL testing conditions, we remark the good performance reported by them for BHandHLYP and the poor performance of the  \gm{Minnesota functionals} family, which is consistent with our findings and those in Ref. \cite{sarkar_benchmarking_2021}}. \gm{A look at the second column of Table 2 shows that the worst-performing functionals are the only ones of the meta-GGA family, including a kinetic energy density correction. This kinetic energy correction is not gauge-invariant and to cope with the presence of the electromagnetic field it should be corrected with the current density \cite{sen_local_2018,sen_benchmarking_2020}. Likely, the absence of gauge invariance is responsible for the poor performance of the Minnesota functionals reported here.}

\section{Conclusions}

A new independent  DFT functional  \tb{assessment} has been proposed, which does not need any external reference. Among the functionals taken into account, \tb{LC-B97D, LC-BP86,} B3P86 \tb{and BHandHLYP} turn out to be  \tb{good }choice\tb{s} for the calculation of properties which require access to electronic excited-state information.

In agreement with a global  \tb{ranking} index I, which summarizes all the mean absolute deviations deduced from off-diagonal hypervirial relationships combined with quantum mechanical sum rules of charge and current conservation, DFT functionals can be separated into three groups according to: i) $0 < \textrm{I} < 2 $, to be used with a good degree of confidence; $ 2 < \textrm{I} < 2.5 $, to be used with some caution; iii) $ \textrm{I} \gg 3  $, not recommended for the calculation of molecular properties resulting from an interaction with external fields. 

Eventually, we would like to emphasize that devising approximate functionals which reduce off-diagonal hypervirial relationship deviations could indicate one of the possible  \tb{ways to} 
 \tb{improve functionals for specific task as the calculation of second order electromagnetic molecular properties.}

\section*{Declaration of Competing Interest}
The authors declare no conflict of interest.

\section*{Acknowledgements}
\gm{The authors are thankful to Prof. K. Hirao for correspondence}. Financial support from the MIUR {(FARB {2018} and FABR {2019})} is gratefully acknowledged. 

\section*{References}

\bibliography{bibfile}

\begin{thebibliography}{10}
\bibitem{nota2}
For clarity, two exceptions were found calculating the TRK sum rule in the
  mixed PR formalism using the MN12-SX functional for ethane and methanol, see
  Table S23; four more exceptions were found for ethane calculating the sum
  rules for the translational invariance of the magnetizability (RL and PL) and
  the deviations relative to polarizability and optical activity using the
  LC-B97D functional, see Tables S25--S28.
\end{thebibliography}

\end{document}